\documentclass[a4paper]{article}
\usepackage[n, advantage, operators,sets, adversary,landau,probability,notions,logic,ff,mm,primitives,events, complexity,asymptotics,keys]{cryptocode}
\usepackage[margin=1.25in]{geometry}
\usepackage{cryptocode}
\usepackage{amsmath}
\usepackage{mathtools}
\usepackage[english]{babel}
\usepackage{hyperref}
\usepackage{pgfplots}
\usepackage{tikz}
\usepackage{url}
\usepackage{braket}
\usepackage{graphicx}
\usepackage[final]{pdfpages}
\hypersetup{
	linktoc=all
}
\DeclarePairedDelimiter{\round}\lfloor\rceil

\title{Intuitive Understanding of Quantum Computation and Post-Quantum Cryptography}
\author{Nguyen Thoi Minh Quan
\footnote{https://www.linkedin.com/in/quan-nguyen-a3209817,
https://scholar.google.com/citations?user=9uUqJ9IAAAAJ, https://github.com/cryptosubtlety, msuntmquan@gmail.com}
\footnote{When asked, "How to say your name, Quan?", I answered, "It's the prefix of Quantum :)".}}
\begin{document}
\date{}
\maketitle
\begin{abstract}
Post-quantum cryptography is inevitable. National Institute of Standards and Technology (NIST) starts standardizing quantum-resistant public-key cryptography (aka post-quantum cryptography). The reason is that investment in quantum computing is blooming which poses significant threats to our currently deployed cryptographic algorithms. As a security engineer, to prepare for the apocalypse in advance, I've been watching the development of quantum computers and post-quantum cryptography closely. Never mind, I simply made up an excuse to study these fascinating scientific fields :) However, they are extremely hard to understand, at least to an amateur like me. This article shares with you my notes with the hope that you will have an \textit{intuitive understanding} of the beautiful and mind-blowing quantum algorithms and post-quantum cryptography.

\textbf{Update}: Multivariate signature scheme Rainbow is broken by Ward Beullens \cite{rainbowbroken}. Supersingular Isogeny Diffie-Hellman protocol (SIDH) is broken by Wouter Castryck and Thomas Decru \cite{sidhbroken}.
\end{abstract}

\tableofcontents
\section*{Introduction}
To understand post-quantum cryptography, we have to understand quantum mechanics and quantum computers. Therefore, before studying post-quantum cryptography, I bought the classic book "Quantum Computation and Computation Information" by Michael Nielsen and Isaac Chuang \cite{quantumbook}. The book was too advanced to me, so I took an excellent approachable quantum computation course by Umesh Vazirani \cite{quantumberkeley}. In fact, I learned most quantum computation from Vazirani's course. Later on, I've realized that understanding quantum computers has nothing to do with understanding post-quantum cryptography. It was too late. I couldn't unlearn what I have learned. Therefore, I'll describe both quantum computers and post-quantum cryptography to make sure that you will make the same mistake as I did :) Joking aside, as a security engineer who is trained with "trust, but verify" mindset, I feel guilty to blind trust that Shor's quantum algorithms \cite{shorquantum} break our current cryptographic protocols.

After "wasting" our time studying quantum algorithms, we'll study post-quantum cryptography. As Shor's algorithms solve factoring and discrete log problems in polynomial time, cryptographers had to find alternative cryptographic constructions that are presumably safe against quantum computers. The following cryptographic constructions are selected to advance to the 2nd round in NIST's post-quantum cryptography competition \cite{nistpostquantum}: lattice-based cryptography, hash-based digital signature, code-based cryptography. I feel a headache just by reading these names :)  They're independent of each other and each topic deserves its own research, so I'll describe them one-by-one in later chapters. They're are all difficult to understand, but the most challenging obstacle is to overcome our fear in dealing with them. No worries, if you can't understand them, blame me for not explaining them well :)

\section{Quantum Computation}
Have you ever played computer games? American player Kyle Giersdorf won 3 million on the Fortnite game. Computer games are ruling the world, so I recommend you stop reading this article, instead go and play games :) Computer games are a strange world where games' creators invent rules and players follow with no questions asked. In the same spirit, we'll follow quantum computers's rules, play along and design quantum algorithms based on its rules. The rules are strange but they're not stranger than computer games' rules. Furthermore, we'll study quantum computation without saying a word about quantum physics. I don't even try to understand quantum physics because I don't want to be crazy :)

\subsection{Quantum computers}
To describe a classical computational system, we define its state, how to change its state and how to measure its state. For instance: 
\begin{itemize}
\item State: $n$ bits $x = x_1,x_2\cdots ,x_n$ represented by $n$ transistors.
\item Classical logic gates such as NOT, AND, OR, NAND, etc are used to change $x_1, x_2, \cdots ,x_n$.
\item Measurement: measure the transistors, based on transistors' voltages, we'll get $n$ output bits.
\end{itemize}

In a similar way, to describe quantum computational system, we'll define quantum state, quantum gates and quantum measurement.

\subsubsection{Quantum state}
In classical computers, a bit is either $0$ or $1$ at any moment. In quantum computers, a quantum bit (aka qubit) can exist at both states $0$ and $1$ at the same time. In fact, a qubit is a superposition of states $\ket{0}$ and $\ket{1}$: $\ket{q} = \alpha_0\ket{0} + \alpha_1\ket{1}$ where the amplitudes $\alpha_0, \alpha_1$ are complex numbers. On the one hand, the $ket$ $\ket{}$ notation $\ket{q}, \ket{0}, \ket{1}$ just means quantum states, instead of classical ones. On the other hand, $\ket{q}$ denotes the state vector $\ket{q} = \begin{pmatrix} \alpha_0 \\ \alpha_1 \end{pmatrix} = \alpha_0\ket{0} + \alpha_1\ket{1}$.

Is it strange that $\alpha_0, \alpha_1$ are complex numbers instead of real numbers? It's even stranger to learn that we never have access to $\alpha_0, \alpha_1$. As we'll see in the later section, when we measure $\ket{q}$, we'll get $\ket{0}$ with probability $|\alpha_0|^2$ and $\ket{1}$ with probability $|\alpha_1|^2$. I.e., we can observe these complex numbers' magnitudes (which are real numbers), but not the numbers themselves. Nature is mysterious!

Generalize the previous paragraphs, 2 qubits is a superposition of states $\ket{00}, \ket{01}, \ket{10}, \ket{11}$, for instance, $1/2(\ket{00} + \ket{01} + \ket{10} + \ket{11})$. $n$ qubits is a superposition of states $\ket{00\cdots0}$, $\cdots$, $\ket{11\cdots1}$, i.e., $\ket{q} = \sum\limits_{x = 0}^{2^n - 1} \alpha_x \ket{x}$. It's amazing that $n$ qubits hold information of \emph{$2^n$} states at the same time. This makes quantum computers more powerful than classical computers.

\subsubsection{Quantum gates}
In classical computers, we use classical logic gates such as AND, NOT, OR, NAND to change bits' values. In quantum computers, we use quantum gates to change qubits. Recall that $\ket{q}$ denotes state vector $\ket{q} = \begin{pmatrix} \alpha_0 \\ \alpha_1 \end{pmatrix} = \alpha_0\ket{0} + \alpha_1\ket{1}$. To transform $2\times 1$ column vectors, we will use $2\times2$ matrix. Therefore, we can describe quantum gates in the form of matrices.

\paragraph{Bit flip gate X}
Bit flip gate $X = \begin{pmatrix} 0 & 1 \\ 1 & 0 \end{pmatrix}$. We have $X\ket{q} = \begin{pmatrix} 0 & 1 \\ 1 & 0 \end{pmatrix}\begin{pmatrix} \alpha_0 \\ \alpha_1 \end{pmatrix} = \begin{pmatrix} \alpha_1 \\ \alpha_0 \end{pmatrix}$, i.e., it transforms $\ket{0}$ into $\ket{1}$ and $\ket{1}$ into $\ket{0}$.

\paragraph{Phase flip gate Z}
Phase flip gate $Z = \begin{pmatrix} 1 & 0 \\ 0 & -1 \end{pmatrix}$. We have $Z\ket{q} = \begin{pmatrix} 1 & 0 \\ 0 & -1 \end{pmatrix}\begin{pmatrix} \alpha_0 \\ \alpha_1 \end{pmatrix} = \begin{pmatrix} \alpha_0 \\ -\alpha_1 \end{pmatrix}$. If we denote $\ket{+} = 1/\sqrt{2}(\ket{0} + \ket{1})$ and $\ket{-} = 1/\sqrt{2}(\ket{0} - \ket{1})$ then we have $Z\ket{+} = \begin{pmatrix} 1 & 0 \\ 0 & -1 \end{pmatrix}\begin{pmatrix} 1/\sqrt{2} \\ 1/\sqrt{2} \end{pmatrix} = \begin{pmatrix} 1/\sqrt{2} \\ -1/\sqrt{2} \end{pmatrix} = \ket{-}$ and $Z\ket{-} = \ket{+}$.

The bit flip gate $X$ and phase flip gate $Z$ are basic and boring as they just transform back and forth between standard states $\ket{0} \leftrightarrow \ket{1}$, $\ket{+} \leftrightarrow \ket{-}$. The following 2 quantum gates are more interesting.
\paragraph{CNOT (Controlled-NOT) gate}
\begin{center}
\includegraphics[scale=0.25]{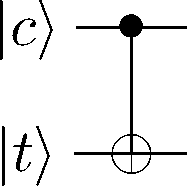}
\end{center}

CNOT (Controlled-NOT) gate: $\ket{c}\ket{t} \rightarrow \ket{c}\ket{c \xor t}$ where $c$ is the control bit, $t$ is the target bit. What it means is that if $c$ is $1$ then it flips the target bit, otherwise it keeps the target bit as is. This can be generalized to implement if/else: if $c$ is $1$, execute gate $U^c = U$, else don't execute $U$ (i.e. execute $U^c = U^0 = I$ identity).
\begin{center}
\includegraphics[scale=0.25]{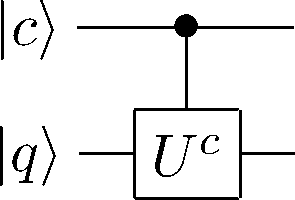}
\end{center}

\paragraph{Hadamard gate H}
Hadamard transform $H = \frac{1}{\sqrt{2}} \begin{pmatrix} 1 & 1 \\ 1 & -1 \end{pmatrix}$. We have $H\ket{0} = 1/\sqrt{2}(\ket{0} + \ket{1})$, $H\ket{1} = 1/\sqrt{2}(\ket{0} - \ket{1})$. If we apply $H^{\bigotimes 2}$ (this notation just means we apply $H$ to 1st and 2nd qubit independently) to 2 qubits $\ket{00}$, we have $H^{\bigotimes 2} \ket{00} = 1/\sqrt{2}(\ket{0} + \ket{1})1/\sqrt{2}(\ket{0} + \ket{1}) = 1/2(\ket{00} + \ket{01} + \ket{10} + \ket{11})$. Observe that Hadamard gate transforms basis state into superposition of states, for instance, H transforms $\ket{0}$ into superposition of states $\ket{0}$, $\ket{1}$ and $H^{\bigotimes 2}$  transforms $\ket{00}$  into superposition of states $\ket{00}, \ket{01}, \ket{10}, \ket{11}$. We have $H^{\bigotimes n} \ket{00\cdots0} = \frac{1}{2^{n/2}}\sum\limits_{x = 0}^{2^n - 1}\ket{x}$, i.e., we can create superposition of all basis states $\ket{0...0}, \cdots, \ket{1...1}$ by applying $H^{\bigotimes n}$ gate to $\ket{0...0}$. We'll use this trick over and over again in designing quantum algorithms.

\begin{center}
\includegraphics[scale=0.25]{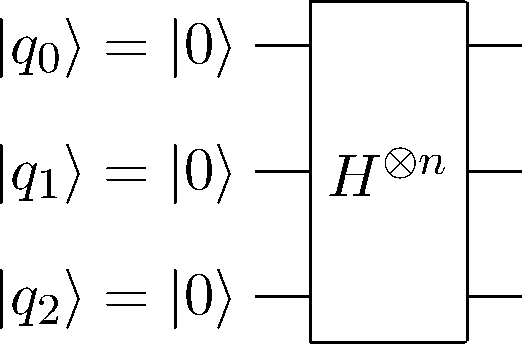}
\end{center}

If we take a closer look at Hadamard gate $H\ket{0} = 1/\sqrt{2}(\ket{0} + \sqrt{2}\ket{1})$, $H\ket{1} = 1/\sqrt{2}(\ket{0} - \ket{1})$, we'll see that $H\ket{u} = 1/\sqrt{2}(\ket{0} + (-1)^{u}\ket{1})$. Therefore
\begin{align*}
H^{\bigotimes 2}\ket{u_1u_2} &=  1/\sqrt{2}(\ket{0} + (-1)^{u_1}\ket{1})1/\sqrt{2}(\ket{0} + (-1)^{u_2}\ket{1}) \\
&= 1/2(\ket{00} + (-1)^{u_2}\ket{01} + (-1)^{u_1}\ket{10} + (-1)^{u_1u_2}\ket{11}) \\
&= 1/2((-1)^{(u_1, u_2).(0,0)}\ket{00} + (-1)^{(u_1,u_2).(0,1)}\ket{01} + (-1)^{(u_1,u_2).(1,0)}\ket{10} + (-1)^{(u_1,u_2).(1,1)}\ket{11}) \\
&= 1/2 \sum_x (-1)^{u.x}\ket{x}
\end{align*}

For $n$ qubits, we have $H^{\bigotimes n} \ket{u_1u_2\cdots u_n} = \frac{1}{2^{n/2}}\sum_x (-1)^{u.x}\ket{x}$ where $u.x = u_1x_1 + u_2x_2 + \cdots + u_nx_n$ is the inner product of $u$, $x$.

\paragraph{Unitary transformation}
If we take a closer look at $X$, $Z$, $H$ we see that $I = X^2 = Z^2 = H^2$. This is not an accident. For every quantum transformation $U$, if we denote $U^*$ as conjugate transpose of $U$ then $UU^* = U^*U = I$. A matrix $U$ satisfying the above equation is called a unitary matrix. In the above examples, it just happens that $X^* = X$, $Z^* = Z$, $H^* = H$ and hence $X^2 = XX^* = I$, $Z^2 = ZZ^* = I$, $H^2 = HH^* = I$.

\subsubsection{Quantum measurement}
In classical computers, measurement is a trivial operation because what you see is what you get. In quantum computers, we never have access to the qubit $q$'s complex amplitudes $\alpha_0, \alpha_1$. When we measure $\ket{q} = \alpha_0\ket{0}  + \alpha_1\ket{1}$, we'll see $\ket{0}$ with probability $|\alpha_0|^2$ and see $\ket{1}$ with probability $|\alpha_1|^2$. Furthermore, the measured qubit forever collapses to state $\ket{0}$ or $\ket{1}$, i.e., you can't never put the measured qubit back to superposition of states $\ket{0}$, $\ket{1}$. To a certain extent, quantum measurement is a destructive operation. Finally, to make sure that the probabilities of measurement outcomes summing up to 1, $\alpha_0$ and $\alpha_1$ must satisfy the equation $|\alpha_0|^2 + |\alpha_1|^2 = 1$. This is the reason why you see $1/\sqrt{2}$ in state like $\ket{+} = 1/\sqrt{2}\ket{0} + 1/\sqrt{2}\ket{1})$ because $(1/\sqrt{2})^2 + (1/\sqrt{2})^2 = 1$. For $n$ qubits $\ket{q} = \sum\limits_{x = 0}^{2^n - 1} \alpha_x \ket{x}$, we have $\sum\limits_{x = 0}^{2^n - 1} |\alpha_x|^2 = 1$.

The way quantum measurement works restricts quantum computers's computation capability. While $n$ qubits $\ket{q} = \sum\limits_{x = 0}^{2^n - 1} \alpha_x \ket{x}$ holds information of $2^n$ basis states at the same time, every time we measure $q$, we only observe a tiny bit of information: a specific state $\ket{i}$ with probability $|\alpha_i|^2$ and even worst, $\ket{q}$ forever collapses to $\ket{i}$. In other words, most $q$'s information is hidden inside inaccessible $\alpha_j, j\in\{0,1\}^n$. This is the reason why quantum computers is not $2^n$ more powerful than classical computers.

\paragraph{Partial measurement}
If we measure 2 qubits like $1/2\ket{00} + 1/2\ket{01} + 1/\sqrt{2}\ket{10}$, we'll see one of states $\ket{00}$, $\ket{01}$, $\ket{10}$. What would happen if we only measure the 1st qubit? If the measured 1st qubit is $\ket{0}$ then the 2 qubits become $1/\sqrt{2}(\ket{00} + \ket{01})$, i.e., superposition of states where the 1st qubit is $\ket{0}$. Note that the amplitudes change from $1/2$ to $1/\sqrt{2}$ to make sure that the sum of new states' probabilities is 1. Similarly, if we measure the 1st qubit  and we see $\ket{1}$ then the 2 qubits become $\ket{10}$ because among three states $\ket{00}, \ket{01}, \ket{10}$ only $\ket{10}$ has the 1st qubit as 1. 
\subsection{Reversible computation}
The fact that any quantum transformation $U$ is unitary has a profound impact on quantum computation. If we start with quantum state $x$ and we apply quantum computation $U$ to it then we can always get $x$ back by appling $U^*$ to $U\ket{x}$. The reason is that $U^*(U\ket{x}) = (U^*U) \ket{x} = I\ket{x} = \ket{x}$, i.e., all quantum computation is reversible. This contrasts with classical computers where there are many one way functions.

The question is if we are given a classical function $f(x)$, can we implement it using reversible quantum computation? The trick is to carry $x$ to the output as well so that we have enough information to reverse the computation. There is quantum circuit $U_f$ that implements $U_f\ket{x}\ket{b} = \ket{x}\ket{b \xor f(x)}$. 

\subsection{Bernstein-Vazirani's algorithm}
To demonstrate the power of quantum computation, we'll take a look at Bernstein-Vazirani's algorithm \cite{parityquantum} that solves parity problems faster than any classical algorithm.

Given a function $f: \{0, 1\}^n \rightarrow \{0,1\}$ as a "black box" \footnote{We can input $x$ to the "black box" and ask it to compute $f(x)$ , but we don't have access to its internal computation process.} and $f(x) = u.x \mod 2$ for some hidden $u \in \{0,1\}^n$, find $u$.

In classical computers, whenever we query the black box $f$ we only get 1 bit of information $f(x)$. As $u \in \{0,1\}^n$ has n bits information, we need at least $n$ queries to the black box $f$ to find n-bit $u$. Furthermore, if we query $f$ n times using the following $x$ values: $10\cdots0$, $010\cdots0$, $\cdots$, $00\cdots1$ then we'll find $u$ because $i^{th}$ query reveals $i^{th}$ bit of $u$. Therefore, the optimal classical algorithm requires $n$ queries to $f$.

In quantum computers, if we denote $U_f$ the quantum circuit that implements $f$ then Bernstein-Vazirani's algorithm only uses $U_f$ once. We'll describe the algorithm by working backwards. Let's recall what we've learned in the previous sections.
\begin{itemize}
\item When we measure n qubits $\ket{q} = \sum\limits_{i = 0}^{2^n - 1} \alpha_i \ket{i}$, we'll see state $\ket{i}, i \in \{0,1\}^n$ with probability $|\alpha_i|^2$. In a special case when $\alpha_u = 1$, $\alpha_{i \neq u} = 0$ or $\ket{q} = \ket{u}$, measuring $\ket{q}$ gives us $\ket{u}$.
\item $H^{\bigotimes n}\ket{u} = H^{\bigotimes n} \ket{u_1u_2\cdots u_n} = \frac{1}{2^{n/2}}\sum_x (-1)^{u.x}\ket{x}$.
\item $H^2 = I$. This implies $H^{\bigotimes n}H^{\bigotimes n}\ket{x} = I^{\bigotimes n} \ket{x} = \ket{x}$.
\end{itemize}

Based on the above facts, if we can set up quantum state $\frac{1}{2^{n/2}}\sum_x (-1)^{u.x}\ket{x} = H^{\bigotimes n}u$ then applying $H^{\bigotimes n}$ to it gives $\ket{u}$, so the final quantum measurement results in $\ket{u}$. Now, the question is how to set up the quantum state $\frac{1}{2^{n/2}}\sum_x (-1)^{u.x}\ket{x}$. Note that we don't know $u$. However, we know $u.x = f(x)$, which means that we can rewrite the above state as $\frac{1}{2^{n/2}}\sum_x (-1)^{f(x)}\ket{x}$. We make progress, but we're not done yet. Even though we don't know how to create state $\frac{1}{2^{n/2}}\sum_x (-1)^{f(x)}\ket{x}$ yet, we know how to create the state that looks similar to it using Hadamard transform $H^{\bigotimes n}\ket{00\cdots0} = \frac{1}{2^{n/2}}\sum_x \ket{x}$. The last bit of the puzzle is how to create $(-1)^{f(x)}$. From "Reversible computation" section, we have $U_f\ket{x}\ket{b} = \ket{x}\ket{b \xor f(x)}$. Let's see what happens to $U_f\ket{x}\ket{-} = 1/\sqrt{2}U_f\ket{x}(\ket{0} -\ket{1}) = 1/\sqrt{2}(U_f\ket{x}\ket{0} - U_f\ket{x}\ket{1}) = 1/\sqrt{2}(\ket{x}\ket{f(x)} - \ket{x}\ket{1 \xor f(x)}) = 1/\sqrt{2}\ket{x}(\ket{f(x)} - \ket{1 \xor f(x)}) = 1/\sqrt{2}\ket{x}(-1)^{f(x)}(\ket{0} - \ket{1}) = \ket{x}(-1)^{f(x)}\ket{-}$. Now, we have $(-1)^{f(x)}$.

\begin{center}
\includegraphics[scale=0.25]{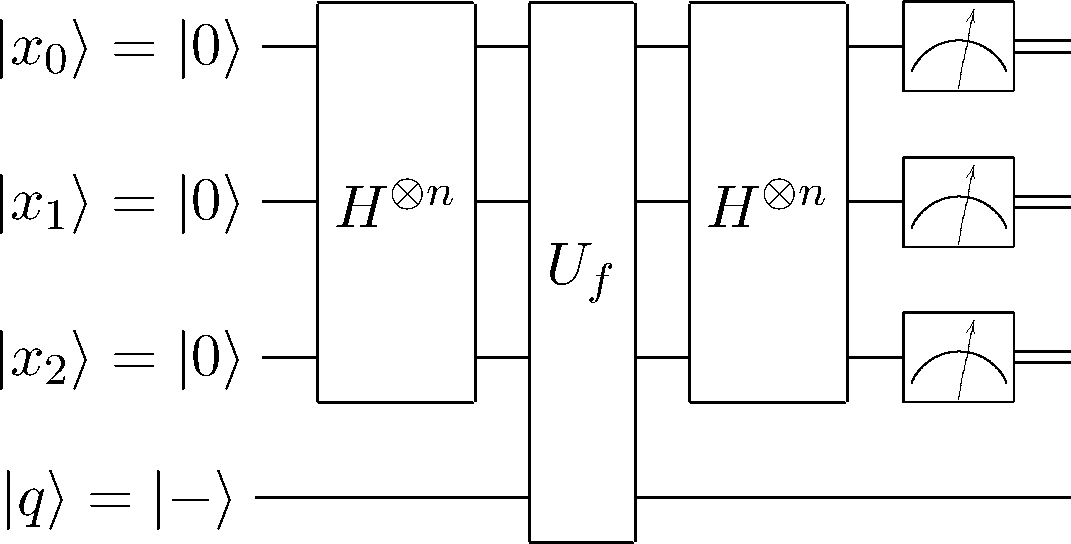}
\end{center}

Putting together in the forward order, we have the algorithm as shown in the above figure:
\begin{enumerate}
\item $H^{\bigotimes n}\ket{00...0}$ gives us $\frac{1}{2^{n/2}}\sum_x \ket{x}$.
\item Applying $U_f$ to $\frac{1}{2^{n/2}}\sum_x \ket{x}\ket{-}$ gives us $\frac{1}{2^{(n/2}}\sum_x (-1)^{f(x)}\ket{x}\ket{-} = \frac{1}{2^{n/2}}\sum_x (-1)^{u.x}\ket{x}\ket{-}$
\item Apply $H^{\bigotimes n}$ to the first n qubits of $\frac{1}{2^{n/2}}\sum_x (-1)^{u.x}\ket{x}\ket{-}$ gives us $\ket{u}\ket{-}$.
\item Measure the first n qubits of $\ket{u}\ket{-}$ gives us $\ket{u}$.
\end{enumerate}

Now, we solved the problem but it's not clear why using $U_f$ once gives us $n$ bit information about $u$. If we look closer at the 2nd step, even though we use $U_f$ once, we see that the sum $\frac{1}{2^{n/2}}\sum_x (-1)^{f(x)}\ket{x}\ket{-}$ have $f(x)$ for all $x\in\{0,1\}^n$. How come? The reason is that $U_f$'s input can be a superposition of all basis states $\sum_x\ket{x}$ (this is in contrast with classical computers where the input is only 1 specific $\ket{i}$) and so the output contains all $f(x), x\in\{0,1\}^n$.

\subsection{Simon's algorithm}
Given a 2-to-1 function $f: \{0, 1\}^n \rightarrow \{0,1\}^n$ and $f(x) = f(x \xor s)$ for some secret $s\in \{0, 1\}^n$, find $s$.

Designing classical algorithms is hard, let alone quantum algorithms. In this problem, the condition $f(x) = f(x \xor s)$ is not even natural, so it's tough to even start the thinking process on how to solve the problem. We have no clue. One method that I found helpful is to play with our existing knowledge, analyze them and/or try to extend them with the hope that somehow it will lead us closer to the solution.

In the previous section, we've seen how to access all $f(x), x\in \{0,1\}^n$ using $H^{\bigotimes n}$, followed by $U_f$. Recall that $H^{\bigotimes n}$ gives us $\frac{1}{2^{n/2}}\sum_x\ket{x}$ and applying $U_f$ to $\frac{1}{2^{n/2}}\sum_x \ket{x}\ket{00...0}$ gives us $\frac{1}{2^{n/2}}\sum_x \ket{x}\ket{f(x)}$. Note that $\frac{1}{2^{n/2}}\sum_x \ket{x}\ket{f(x)}$ are 2n qubits as $f(x)$ is n-bit.

If we measure the 1st n qubits of $\frac{1}{2^{n/2}}\sum_x \ket{x}\ket{f(x)}$ and the 1st n qubits collapses to $\ket{r}$ for some $r$, then the 2n qubits collapses to $\ket{r}\ket{f(r)}$. We've made progress as we've just learned that this path leads nowhere :)

\begin{center}
\includegraphics[scale=0.25]{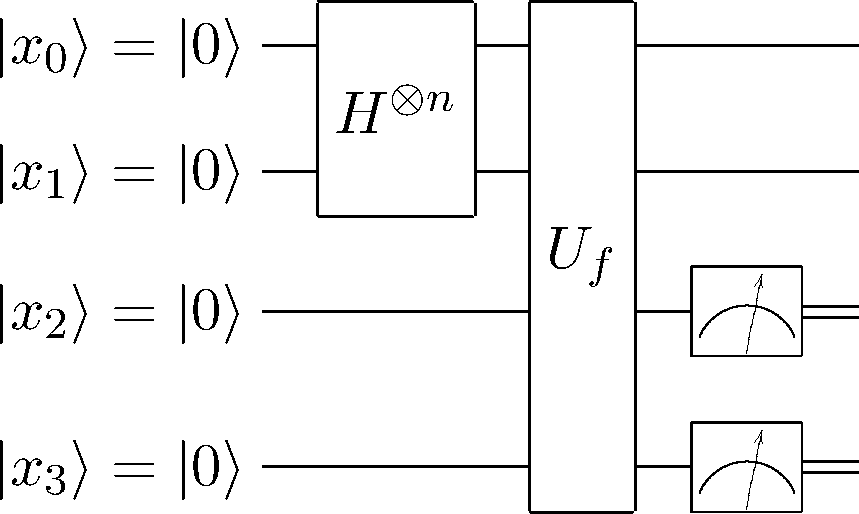}
\end{center}

If we measure the 2nd n qubits of $\frac{1}{2^{n/2}}\sum_x \ket{x}\ket{f(x)}$ and we see $\ket{f(r)}$ for some $r$ then the 2n qubit collapses to $\ket{r}\ket{f(r)}$. Wait a second, this wasn't correct. From the problem statement, we know $\ket{f(r)} = \ket{f(r \xor s)}$, i.e., if we see $\ket{f(r)}$, then we see $\ket{f(r \xor s)}$ as well because they both equal each other. In other words, the 2n qubits state collapses to $1/\sqrt{2}(\ket{r}\ket{f(r)} + \ket{r \xor s}\ket{f(r \xor s)}) = 1/\sqrt{2}((\ket{r} + \ket{r \xor s})\ket{f(r)})$. This is pretty cool as we've discovered a new state $1/\sqrt{2}(\ket{r} + \ket{r \xor s})$ that we've never seen before.

What are we going to do with $1/\sqrt{2}(\ket{r} + \ket{r \xor s})$? This is a specific state for some random $r$, so let's apply $H^{\bigotimes n}$ to make it in superposition of all basis states again. Are you sick of $H^{\bigotimes n}$ yet? :) I'm sorry that I have to make you real sick because we'll continue using $H^{\bigotimes n}$ over and over again. Applying $H^{\bigotimes n}$ to $\ket{r}$ gives us $\frac{1}{2^{n/2}}\sum_x (-1)^{r.x}\ket{x}$ while applying $H^{\bigotimes n}$ to $\ket{r \xor s}$ gives us $\frac{1}{2^{n/2}}\sum_x (-1)^{(r \xor s).x}\ket{x}$, therefore applying $H^{\bigotimes n}$ to $1/\sqrt{2}(\ket{r} + \ket{r \xor s})$ gives us $\frac{1}{2^{(n + 1)/2}}\sum_x ((-1)^{r.x} + (-1)^{(r \xor s).x})\ket{x}$. The value $(-1)^{r.x} + (-1)^{r.x \xor s.x}$ is pretty special because if $s.x = 1 \mod 2$ then $(-1)^{r.x} + (-1)^{r.x \xor 1} = (-1)^{r.x} (1 + (-1)) = 0$ (i.e. we never see the corresponding $\ket{x}$), otherwise if $s.x = 0$ then $(-1)^{r.x} + (-1)^{r.x \xor 0} = 2(-1)^{r.x} \neq 0$. What it means is after measurement, we only see $\ket{x}$ such that $s.x = 0$ for some random $x$. The complete Simon algorithm \cite{simonquantum} is shown in the below figure

\begin{center}
\includegraphics[scale=0.25]{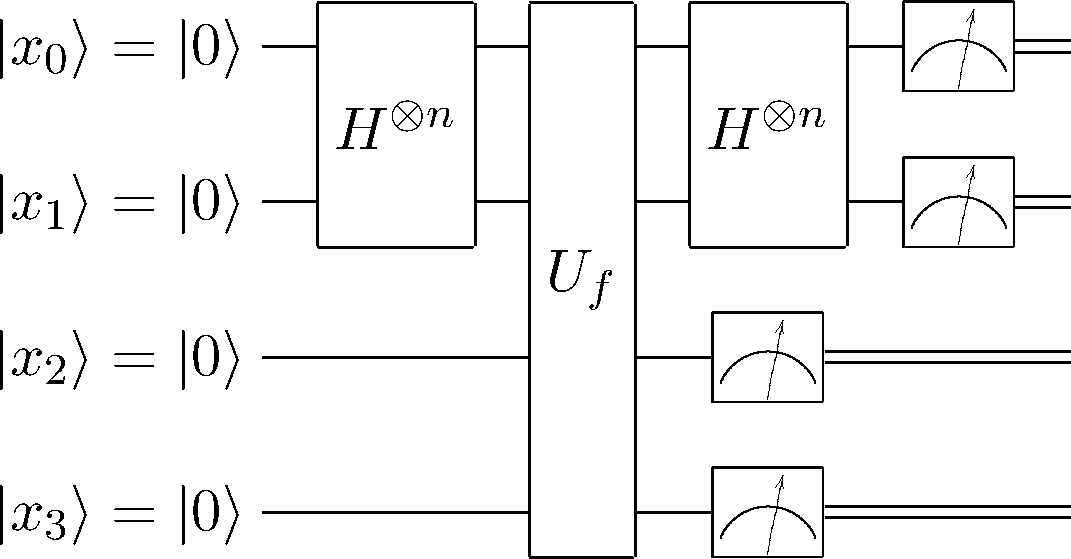}
\end{center}

All right, we get $s.x = 0$ for some random $x$. This isn't enough to solve for $s$. What we can do is to repeat the algorithm so that we will receive a system of linear equations where each equation has the form $s.x = 0$ with probably different random $x$. Roughly we need $n$ equations of the form $s.x = 0$ to find $s$ using Gauss elimination algorithm.

\subsection{Quantum Fourier Transform (QFT)}
Denote $\omega$ N-rooth of unity: $\omega^N = 1$. Classical Fourier transform is defined by the following matrix:

$F_N = 1/\sqrt{N}
\begin{pmatrix}
1 & 1 & 1 & \cdots & 1 \\
1 & \omega & \omega^2 & \cdots & \omega^{N - 1} \\
1 & \omega^2 & \omega^4 & \cdots & \omega^{2(N - 1)} \\
\vdots & \vdots & \vdots & \ddots & \vdots \\
1 & \omega^{N - 1} & \omega^{2(N - 1)} & \cdots & \omega^{(N - 1)^2}
\end{pmatrix}$

Quantum Fourier Transform (QFT) is simply applying the same matrix to a quantum state: $F_N \begin{pmatrix} \alpha_0 \\ \alpha_1 \\ \vdots \\ \alpha_{N - 1} \end{pmatrix} = \begin{pmatrix} \beta_0 \\ \beta_1 \\ \vdots \\ \beta_{N - 1} \end{pmatrix}$ or $F_N (\sum_j \alpha_j\ket{j}) = \sum_k \beta_k\ket{k}$. In a special case, when $\alpha_0 = 1$ and $\alpha_i = 0, i \neq 0$ we have $F_N\ket{0} = 1/\sqrt{N} \sum_k\ket{k}$, i.e., it transform $\ket{0}$ into superposition of all basis states. Did you notice that it looks similar to Hadamard transform $H^{\bigotimes n}$? The reason is that in fact, Hamamard transform is a special case of QFT.

I don't know about you but $F_N$ looks scary as hell to me :) Fourier transform is beyond my depth, let alone QFT. However, QFT plays an important role in Shor's algorithms, so we have to study it. One method to get away from dealing with deep math is to learn its properties without proving them. We can dig deeper into the proof once our math muscle is stronger. For now, let's see QFT's nice properties.

\paragraph{Efficient implementation} QFT can be implemented in $O((logN)^2)$. In other words, QFT can be efficiently implemented in polynomial time of N's bit length.

\paragraph{Shift property} $F_N \begin{pmatrix} \alpha_{N - 1} \\ \alpha_0 \\ \alpha_1 \\ \vdots \\ \alpha_{N - 2} \end{pmatrix} = \begin{pmatrix} 1 \beta_0 \\ \omega \beta_1 \\ \omega^2 \beta_2 \\ \vdots \\ \omega^{N - 1} \beta_{N - 1} \end{pmatrix}$ or $F_N (\alpha_{N _ 1}\ket{0} + \alpha_0\ket{1} + \alpha_1\ket{2} + \cdots + \alpha_{N - 2} \ket{N - 1} = \sum_k \omega^k\beta_k\ket{k}$, i.e., when we move the 1st amplitude $\alpha_0$ to the 2nd position, the 2nd amplitude $\alpha_1$ to the 3rd position, etc then applying QFT to the result corresponds to multiplying $\beta_k$ with $\omega^k$. When we measure $\sum_k \omega^k\beta_k\ket{k}$, we'll see specific state $\ket{k}$ with probability $|\omega^k\beta_k|^2 = {|\omega^k|}^2|\beta_k|^2$. As the magnitude of $\omega^k$ is 1, we'll see specific state $\ket{k}$ with probability $|\beta_k|^2$. This is exactly the same as if we measure $\sum_k\beta_k\ket{k}$. What it means is that if we shift the input $\alpha_i$, it doesn't change the measurement of the QFT's output.

\paragraph{Periodic property} If $f(x)$ is a periodic function with period $r$ (i.e., $f(x) = f(x + r \mod N)$) then $F_Nf(x)$ is a periodic function with period $N/r$. Let's consider a special case $f(x) = 1/\sqrt{N/r}\sum_{j = 0}^{N/r - 1} \ket{jr}$, i.e., the amplitudes are  $1/\sqrt{N/r}$ at $0, r, 2r, \cdots, (N/r - 1)r$ and are zeros everywhere else, hence $f(x)$ is a period function with period $r$. In this special case, $F_N(1/\sqrt{N/r}\sum_{j = 0}^{N/r - 1} \ket{jr}) = 1/\sqrt{r}\sum_{l = 0}^{r - 1} \ket{lN/r}$. Note that the right hand side's amplitudes are $1/\sqrt{r}$ at $0, N/r, 2(N/r)$, $\cdots$, $(r - 1)(N/r)$ and are zeros everywhere else and hence it's a period function with period $N/r$.

\subsection{Period finding}
Assume $f: \{0, 1, \cdots, N - 1\} \rightarrow S$ ($S$ is just some set) is a periodic function with hidden period $r$ ($r | N$): $f(x) = f(x + r (\mod N))$. Find $r$.

The algorithm is shown in the left figure.

\begin{figure}[h]
\includegraphics[scale=0.18]{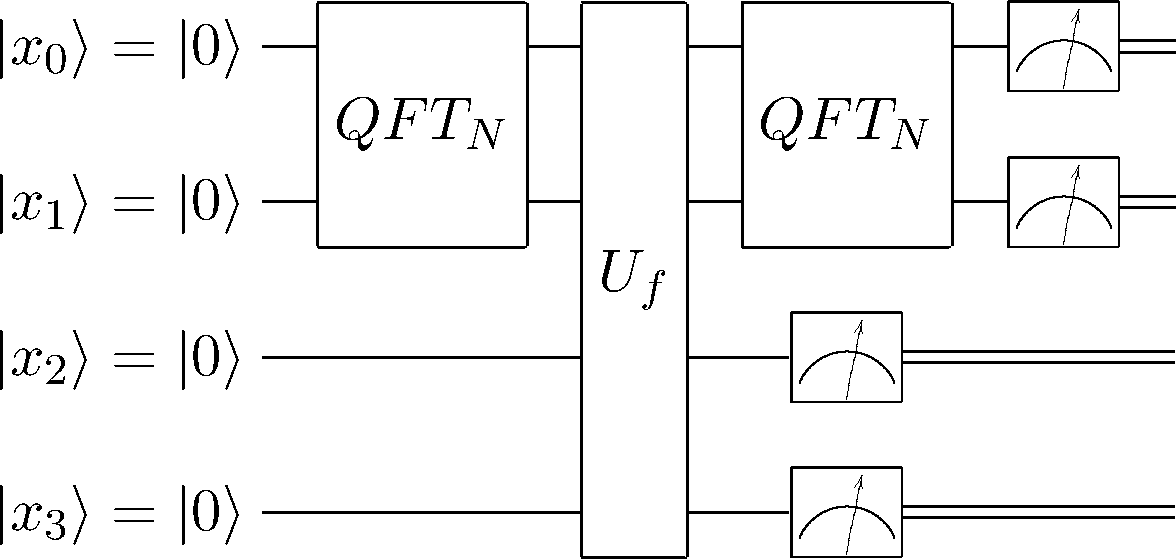} \hspace*{1cm} \includegraphics[scale=0.18]{simon.png}
\end{figure}

I guess you're disappointed with me because I didn't explain the thinking process in designing the solution, instead I jumped directly to the final algorithm as shown above. The reason is that if you look back Simon's algorithm (as shown in the right figure), you'll notice that it looks exactly the same as period finding's algorithm, except we replace $H^{\bigotimes n}$ in Simon's algorithm with $QFT_N$ in period finding algorithm. Surprise!

The function $f$ in period finding problem is different from Simon's algorithm and $QFT_N$ is similar but different from $H^{\bigotimes n}$. Therefore, we have to analyze the period finding algorithm closely.

After applying $QFT_N$ to $\ket{0}$, we get $1/\sqrt{N}\sum_x \ket{x}$. After $U_f$, we have $1/\sqrt{N}\sum_x \ket{x}\ket{f(x)}$. If we measure $\ket{f(x)}$ and get $f(k)$ for some specific $k$ then as $f(k) = f(k + r) = \cdots = f(k + r(N/r - 1))$, we know that $1/\sqrt{N}\sum_x \ket{x}\ket{f(x)}$ collapses to $\ket{k}\ket{f(k)}  + \ket{k + r}\ket{f(k + r)}$ + $\cdots$ + $\ket{k + r(N/r - 1)} \ket{f(k + r(N/r - 1))}$. Rewrite the result as $\sum_{j = 0}^{N/r - 1} \ket{jr + k}\ket{f(k)}$. However, as shifting doesn't affect QFT's measurement, we can simplify the 1st n qubits to $\sum_{j = 0}^{N/r - 1} \ket{jr}$. Now, using period property of QFT, we know that the last $QFT_N$ will transform $\sum_{j = 0}^{N/r - 1} \ket{jr}$ into  $\sum_{l = 0}^{r - 1} \ket{lN/r}$. After measuring the 1st n qubits, we'll get $sN/r$ for some random $s$.

If we repeat the above algorithm a few times, we get $s_1N/r$, $s_2N/r$, $s_3N/r$, etc. It's easy to see that the greatest common divisor of them gives us $N/r$ or $r$.

\subsection{Shor's algorithms}
In the previous section, we have built a general framework to find the period of an arbitrary periodic function. In this section, we'll describe Shor's algorithms \cite{shorquantum} by transforming factoring problems and discrete log problems into period finding problems. 
\subsubsection{Shor's factoring algorithm}
Given $n = p_1*p_2$ where $p_1, p_2$ are prime numbers, find $p_1, p_2$. 

Shor's algorithm is based on the following classical observation. For a random number $x$, if we can find an even number $r$ such that $x^r = 1 \mod n$ then $gcd(x^{r/2} - 1, n)$ or $gcd(x^{r/2} + 1, n)$ gives us the factor of $n$. Let's see why. $x^r - 1 = (x^{r/2} - 1)(x^{r/2} + 1) = 0 \mod n$ implies $(x^{r/2} - 1)(x^{r/2} + 1) = 0 \mod p_1$, i.e., $p_1$ divides either $(x^{r/2} - 1)$ or $(x^{r/2} + 1)$. In other words, $p_1$ divides $gcd(x^{r/2} - 1, n)$ or $gcd(x^{r/2} + 1, n)$.

What does $x^r = 1 \mod n$ mean? It means that if we fix the random number $x$ and define $f(a) = x^a \mod n$ then $f(a + r) = x^{a + r} = x^ax^r = x^a*1 = f(a) \mod n$, i.e., $f(a)$ is a periodic function with period $r$. Therefore, the algorithm is to generate random $x$, use the period finding algorithm to find period $r$ of $f(a) = x^a \mod n$ and finally compute $gcd(x^{r/2} - 1, n)$ or $gcd(x^{r/2} + 1, n)$. What if we find odd $r$, instead of even $r$? We just repeat the algorithm with a different $x$.

\subsubsection{Shor's discrete algorithm}
Assuming $p$ is a prime number and $g$ is the generator of multiplicative group $\mod p$, i.e., the set $\{g^0, g^1, \cdots, g^{p - 2}\}$ is the same as $\{1, 2, \cdots, p - 1\}$. Given $y = g^x \mod p$, find $x$.

Let's take a look at the function $f(a, b) = g^{a}y^{-b} \mod p$, we have $f(a + rx, b + r) = g^{a + rx}y^{-b - r} = g^{a + rx}(g^x)^{-b - r} = g^{a + rx - xb - rx} = g^{a - xb} = f(a, b) \mod p$. I.e., $f(a, b)$ is a periodic function with tuple period $(rx, r)$. Using a period finding algorithm to find the period $(r_1, r_2)$ of $f(a, b)$ and compute discrete log $x = r_1/r_2$.

\section{Lattice-based cryptography}

Lattice \cite{winterschoollattice}, \cite{latticecrypto} is a rare double-edged sword in cryptography. It can be used to build cryptographic protocols as well as to break them. Furthermore, lattice-based cryptography is assumed to be safe against quantum computers, while ECDH key exchange and RSA aren't. As quantum computers are more powerful than classical computers, can we deduce that lattice-based cryptography is safe against classical computers? No, we can't. It's because we start with a security assumption (not fact) which might turn out to be wrong. Like many other things in cryptography, lattice-based cryptography's security is an assumption, no one knows for sure. There is no evidence that lattice-based cryptography is safer than ECDH against classical computers. Therefore, it's better to deploy lattice-based key exchange in hybrid mode together with ECDH where the shared key is derived from both lattice and ECDH. Finally, lattice-based cryptography has extraordinary properties that no other cryptographic constructions have. It is used to build famous fully homomorphic encryption (FHE).\footnote{I copied some paragraphs from my previous article \cite{advancedcrypto} because I'm lazy :)}

Lattice-based cryptography often relies on the hardness of finding something small or solving equations with errors. Every time you see the words small and errors in the context of vectors, equations, polynomials, you know that you land in the realms of lattice-based cryptography.

Before studying lattice, let's warm up by solving Fermat's last equation $x^n + y^n = z^n$. I found a beautiful solution $z = \sqrt[n]{x^n + y^n}$. Sorry, I forgot that $x, y, z$ must be integers :) All right, I found an integer solution as well $x = 0, z = y$ :) What's wrong with me! I missed the conditions $x, y, z > 0$. The lessons learned are to pay attention to both the variables' domains and their constraints.

\subsection{Lattice definitions}

\paragraph{Definition} Lattice \footnote{Lattice has a deep theory which is out of my depth. If you want to understand lattice in depth, I recommend studying it by excellent formal sources \cite{winterschoollattice}, \cite{latticecrypto}.} is a set of points $L = \{\sum\limits_{i = 1}^{i = n}a_iv_i| a_i \in \mathbb{Z}\}$ where $v_i$ are linearly independent vectors in $\mathbb{R}^n$ and $B = \{v_1, \cdots, v_n\}$ is a base of $L$. We often write $v_i$ in the form of $n \times 1$ column vector and $B = (v_1, \cdots, v_n)$ is a $n \times n$ matrix. 

To understand this abstract definition, let's take a look at a concrete example where $v_1 = (2, 0), v_2 = (0, 2)$, i.e., $v_1$ and $v_2$ are twice unit vectors in x-axis and y-axis. The lattice $L$ is the set of points $\{a_1v_1 + a_2v_2 = (2a_1, 2a_2)|a_1, a_2 \in \mathbb{Z}\}$, i.e., all even integer points in 2-dimensional space. This gives us a figure of lattice, but we haven't gone far in understanding the definition. One way to move forward is to question the details of the definition.

What if $\{v_1, \cdots, v_n\}$ are dependent vectors, e.g., $v_1 = v_2 = (0, 2)$? In this case, a single vector $v_1$ is enough to generate $L$, i.e., the linearly independent condition is to remove redundancy in $\{v_1, \cdots, v_n\}$.

What if $a_i$ are real numbers, instead of integers? In this case, $L$ is the whole 2-dimensional space regardless of the base $B = \{v_1, v_2\}$ , i.e., $L$ loses all its interesting math structure.

If we change $\{v_1, \cdots, v_n\}$, will $L$ change? Not always. $L$ is defined as a set of points, i.e., the points are the essence of $L$, not vectors $\{v_1, \cdots, v_n\}$ that are used to generate $L$. The vectors $\{v_1'=(2,0), v_2' = (2, 2)\}$ ($v_1' = v_1, v_2' = v_1 + v_2$) also generate all even integer points in 2-dimensional space, i.e., the base $B = \{v_1, \cdots, v_n\}$ of $L$ is not unique. In fact, if $U$ is an integer matrix with determinant $\pm 1$ then $B' = BU$ is another base of $L$. In our example, $U = \begin{pmatrix} 1 & 1 \\ 0 & 1 \end{pmatrix}$, $det(U) = 1$ and $(v_1', v_2') = (v_1 , v_1 + v_2) = (v_1 , v_2) \begin{pmatrix} 1 & 1 \\ 0 & 1 \end{pmatrix}$.

We've made significant progress but we can go further by playing with the elements in the definition. If $a_i = 0$ then $0 = 0v_1 + \cdots + 0v_n$ is in $L$. If $x = a_1v_1 + \cdots a_nv_n$ is in $L$ then $-x = (-a_1)v_1 + \cdots + (-a_n)v_n$ is in $L$. If $x = a_1v_1 + \cdots + a_nv_n$ and $x' = a_1'v_1 + \cdots + a_n'v_n$ are in $L$ then $x + x' = (a_1 + a_1')v_1 + \cdots + (a_n + a_n') v_n$ is also in $L$. The above properties make $L$ an additive subgroup of $\mathbb{R}^n$. The question is whether any additive subgroup of $\mathbb{R}^n$ a lattice? When showing the above properties we haven't used the fact that $a_i$ are integers at all. As mentioned above if $a_i$ are real numbers then $L$ contains all points in 2-dimensional space and the fact that $a_i$ are integers make the lattice look discrete and not dense. Put everything together, we have an equivalent definition of lattice.

\paragraph{Definition} Lattice is a discrete additive subgroup of $\mathbb{R}^n$.

You may wonder why I made an effort to introduce the 2nd definition. What's the point? In practical problems, no one is going to tell us that there is a lattice structure. We ourselves have to recognize the lattice structure hidden in the problem we're trying to solve. The 1st definition is not useful in this regard because we're not even sure yet there is a lattice, let alone know its basis vectors. The 2nd definition gives us signals to detect hidden lattice structure.

Let's look at the set of integer solutions of the following equation $2x_1 + 3x_2 = 0$. It's obvious that $(0, 0)$ is a solution. If $(x_1, x_2)$ and $(x_1', x_2')$ are solutions then $(-x_1, -x_2)$ and $(x_1 + x_1', x_2 + x_2')$ are solutions because $2(-x_1) + 3(-x_2) = -(2x_1 + 3x_2) = 0$ and $2(x_1 + x_1') + 3(x_2 + x_2') = (2x_1 + 3x_2)  + (2x_1' + 3x_2') = 0 + 0 = 0$. The above properties make the set of solutions additive subgroup of $\mathbb{R}^n$ and the integer condition makes it discrete. Therefore, the set of solutions is a lattice. In conclusion, we know that the set of solutions forms a lattice without solving the equation, is it amazing? While this example is somewhat trivial, we'll use this observation in more complicated problems in later sections.

\subsubsection{Successive minima}
We briefly introduce a few concepts that we'll need in later sections.

$\lambda_1(L)$ denotes the length of a nonzero shortest vector in $L$. As $0$ is always in $L$, in the definition of $\lambda_1(L)$, we have to add the nonzero condition to avoid trivial cases.

The $i$th successive minima $\lambda_i(L)$ is the smallest $r$ such that $L$ has $i$ linearly independent vectors whose lengths are at most $r$.

\subsection{Hard lattice computational problems}

Public-key cryptography is based on hard computational problems. For instance, ECDH is based on hardness of discrete log problems while RSA is based on hardness of factoring problems. Similarly, lattice-based cryptography is based on hard lattice computational problems. Therefore, we briefly introduce the following hard lattice computational problems, mostly to introduce the terminologies.

Shortest Independent Vectors Problem (SIVP$_\gamma$): Given lattice $L(B)$ and small $\gamma > 0$, find $n$ linearly independent lattice vectors whose length is at most $\gamma\lambda_n(L)$.

Bounded Distance Decoding Problems (BDD$_\gamma$): Given a lattice $L(B)$, small $\gamma > 0$ and a target point $t$ that is guaranteed to be close to $L$ (i.e. distance$(t, L) < d = \lambda_1(L)/\gamma$), find a unique lattice vector $v$ such that $||v - t|| < d$.

For the purpose of this article, it's fine to just remember the following: for lattice, it's difficult to find short lattice vectors or to find a lattice vector close to a given target point.

\subsection{Short Integer Solution (SIS)}
\paragraph {SIS Problem} Given $m$ random vector $a_i \in \mathbb{Z}_q^n$ (i.e. $a_i$ is a n-dimensional vector where each coordinate is integer $\mod q$), find nonzero solution $(z_1, \cdots, z_m) \in \{-1, 0, 1\}$ of system of linear equations $z_1a_1 + \cdots + z_ma_m = 0$. In matrix form $Az = 0$ where $A = (a_1, \cdots, a_m)$ is a $n \times m$ matrix.

Note that if there is no condition $z_i \in \{-1, 0, 1\}$ then we know how to solve this system of linear equations using Gaussian elimination.

Let's take a look at the set of solutions $\{(z_1, \cdots, z_m)\}$. $(0, \cdots, 0)$ is a trivial solution. If $(z_1, \cdots, z_m)$ and $(z_1', \cdots, z_m')$ are solutions then $(-z_1, \cdots, -z_m)$ and $(z_1 + z_1', \cdots, z_m + z_m')$ are solutions because $(-z_1)a_1 + \cdots + (-z_m)a_m = -(z_1a_1 + \cdots + z_ma_m) = 0$ and $(z_1 + z_1')a_1 + \cdots + (z_m + z_m')a_m = (z_1a_1 + \cdots + z_ma_m) + (z_1'a_1 + \cdots + z_m'a_m) = 0 + 0 = 0$. The above properties make the set of solutions $\{(z_1, \cdots, z_m\}$ an additive group and integer condition makes it discrete. Therefore, the set of integer solutions $\{(z_1, \cdots, z_m)\}$ is a lattice. How's about the condition $z_i \in \{-1, 0, 1\}$? This condition forces the vector $(z_1, \cdots, z_m)$ to be small. Consequently, solving SIS problems is similar to finding small vectors in the lattice of integer solutions. In other words, solving SIS is a hard problem. It's pretty cool as whenever we see a hard problem, we have hope to use it to build cryptographic protocols. In the next section, we'll use the hardness of SIS problem to construct collision-resistant hash function.

\subsubsection{Collision-resistant hash functions based on SIS}

Given a random matrix $A \in Z_q^{n \times m}$, the function $f_A(z) = Az$ where $z \in \{0, 1\}^m$ is a collision-resistant hash function.

We'll use proof by contradiction. If $f_A(z) = Az$ is not a collision-resistant hash function then we can find $z_1, z_2 \in \{0, 1\}^m$ such that $f_A(z_1) = f_A(z_2)$. This implies $Az_1 = Az_2$ or $A(z_1 - z_2) = 0$. As $z_1, z_2 \in \{0, 1\}^m$, we have $z_1 - z_2 \in \{-1, 0, 1\}^m$. It means that we found $z = z_1 - z_2 \in \{-1, 0, 1\}^m$ such that $Az = 0$. In other words, we can solve SIS problems (contradicts the fact that SIS problems are hard). 

\subsection{Learning With Errors (LWE)}

Who came up with the name "learning with errors"\cite{regevlwe}? Learning with accurate information is hard, let alone learning with misinformation and errors :) Maybe, it's the intention to make the attacker's life miserable.

\paragraph{LWE Problem} Generate $m$ random vectors $a_i \in Z_q^n$, small random errors $e_i$, a random secret $s \in Z_q^n$ and computes $b_i = \langle a_i, s\rangle  +  e_i \mod q \in Z_q$ \footnote{The notation $\langle x, y\rangle$ denotes the inner product of 2 vectors $x = (x_1, \cdots, x_n), y = (y_1, \cdots, y_n)$, i.e., $x_1y_1 + \cdots + x_ny_n$}. Given $(a_i, b_i)$ , find $s$.\\
In the matrix form, given $(A, b^t = s^tA + e^t \mod q)$ \footnote{The notation $s^t$ means the transpose of $s$}, find $s$ where $A = (a_1 \cdots a_m)$ is an $n \times m$ matrix and $b = (b_1 \cdots b_m) \in Z_q^m$. A closely related problem is to not find $s$, but to distinguish $(A, b^t = s^tA + e^t \mod q)$ from random distribution.

Note that if there are no errors $e_i$ then $b^t = s^tA$ is a system of linear equations which we know how to solve for $s$ using Gaussian elimination.

Now, let's try to find our lattice in this LWE problem. If we remove errors $e_i$ then if we fix matrix $A$, the set $L = \{s^tA \mod q\}$ forms a lattice. Why? $0 = 0A$ is in $L$. If $x_1 = s_1^tA$ and $x_2 = s_2^tA$ are in $L$ then $-x_1$ and $x_1 + x_2$ are in $L$ because $-x_1 = -s_1^tA = (-s_1^t)A$ and $x_1 + x_2 = s_1^tA + s_2^tA = (s_1^t + s_2^t)A$. The above properties make $L$ additive group and the integer condition makes it discrete. Therefore, $L$ is a lattice. In our LWE problem, $b^t= s^tA + e^t$ is a point close to the lattice point $s^tA$ because the difference between them is the small error vector $e^t$. Our task is given $b^t$, find the lattice point $s^tA$ close to it (note that knowing $s^tA$ is enough to find $s^t$ using Gaussian elimination). This is the Bounded Distance Decoding (BDD$_\gamma$) problem which is hard to solve.

\subsubsection{Ring-LWE}

Define $R = Z_q[x]/(x^n + 1)$ where $n$ is a power of 2, i.e., polynomials of degree $n$ where coefficients are integers $\mod q$ and operations on polynomials are $\mod x^n + 1$ (i.e. we can replace $x^n$ with -1). The definition of Ring-LWE is similar to LWE, except instead of using $Z_q^n$, we use $R$.

\paragraph{Ring-LWE Problem} Generate $m$ random polynomials $a_i \in R$, small random error polynomials $e_i$, a random secret $s \in R$, computes $b_i = a_is + e_i$. Given $(a_i, b_i)$, find $s$.

What's the relationship between element of $R$ with our familiar vector $Z_q^n$? Let's write down a polynomial $p(x)$ in $R$
\begin{flalign*}
p(x) = a_{n - 1}x^{n - 1} + \cdots + a_1x + a_0
\end{flalign*}

The coefficients $(a_{n - 1}, \cdots, a_1, a_0)$ is a vector in $Z_q^n$. One one hand, to a certain extent, we can cast Ring-LWE problems into LWE problems. On the other hand, polynomial $\mod (x^n + 1)$ has more structure than $Z_q^n$. Let's briefly take a look at $p(x)x \mod x^n + 1$
\begin{flalign*}
p(x)x &= (a_{n - 1}x^{n - 1} + a_{n - 2}x^{n - 2} + \cdots + a_1x + a_0)x \\
&= a_{n - 1}x^n + a_{n - 2}x^{n - 1} + \cdots + a_1x^2 + a_0x \\
&= a_{n - 1}(-1) + a_{n - 2}x^{n - 1} + \cdots + a_1x^2 + a_0x \\
& = a_{n - 2}x^{n - 1} + \cdots + a_1x^2 + a_0x - a_{n - 1}
\end{flalign*}

I.e. multiply $p(x)$ with $x$ corresponds to transform from $(a_{n - 1}, \cdots, a_1, a_0)$ to $(a_{n - 2}, \cdots, a_1, a_0, -a_{n - 1})$. Properties like this make implementation of Ring-LWE more efficient than LWE. However, whenever you have extra math structure, it might later come back and help cryptanalysis. There is an on-going debate about the trade-off between efficiency of Ring-LWE and its extra math structure to security. 

\subsection{Regev's LWE public key cryptosystem}

In this section, we'll take a look at Regev's public key cryptosystem \cite{regevlwe} that is based on the hardness of LWE problem. The protocol only encrypts a single bit $\mu$ but it can be generalized to encrypt arbitrary data. Note that Regev's encryption is only semantically secure, i.e., the ciphertext is indistinguishable from random distribution and it's safe against eavesdropper. It's not safe against active adversary, i.e., it's not safe against chosen ciphertext attack.

To build any public key cryptosystem, the first step is to construct private/public key pair. LWE's description gives us a hint on how to do it: Alice's private key is $s$ and her public key is $(A, b^t = s^tA + e^t)$. As $(A, b^t = s^tA + e^t)$ is indistinguishable from random distribution, the public key $(A, b^t)$ doesn't leak any information about the private key $s$. The question is how to use the public key $(A, b^t)$ for encryption?

As semantic security concerns indistinguishability from random, we'll try to generate random alike traffic to confuse eavesdroppers. We need another math property: if $x \in \{0, 1\}^m$ then $(A, Ax)$ is indistinguishable from random distribution. Therefore, using Alice's public key $(A, b^t = s^tA + e^t)$, Bob can generate $x \in \{0, 1\}^m$ and send $(Ax, b^tx)$ to Alice. This is safe because $(Ax, b^tx)$ is indistinguishable from random distribution. What can Alice do? Note that $b^tx = (s^tA + e^t)x = s^tAx + e^tx$ and Alice knows the private key $s^t$, so she can compute $b^tx - s^tAx = e^tx$. I.e., Alice can compute the noise $e^tx$ ($e^t$ and $x$ are small), but the eavesdropper can't. Awesome, Bob can generate random-alike traffic $(Ax, b^tx)$ and Alice can denoise it. We're happy, we make the eavesdropper confused, except we confuse ourselves as well because we haven't encrypted anything yet :)

We're close to the final protocol. The last piece of the puzzle is how to include the bit $\mu$ into the traffic. What will happen if Bob sends $(Ax, b^tx + \mu)$? Alice uses the same computation to get $b^tx - s^tAx = e^tx + \mu$. However, Alice doesn't know $x$, so while she can compute the noise $e^tx + \mu$, she can't extract $\mu$ from it. The idea to solve this puzzle is to encode the "meaning" of $\mu$, instead of $\mu$ itself into the traffic: if $\mu = 0$, we doesn't include it into the traffic, but if $\mu = 1$ then we include large number $\mu\round{\dfrac{q}{2}}$ into traffic. When Alice receives it, she computes $b^tx - s^tAx$ and if the result is small, she knows that $\mu$ was 0, otherwise if $b^tx - s^tAx$ is large, she knows that $\mu$ was 1. The complete protocol is shown in the below figure.
\begin{center}
\pseudocode{%
\textbf{Alice} \<\< \textbf{Bob} \\
\text{Secret s} \sample Z_q^n  \<\<  x \sample \{0, 1\}^m  \\
b^t = s^tA + e^t \< \sendmessageright*{(A, b^t)} \< \\
\<\< u = Ax \\
\<\< u' = b^tx + \mu\round{\dfrac{q}{2}} \\
\< \sendmessageleft*{(u, u')} \\
u' - s^tu \approx \mu\round{\dfrac{q}{2}} \< \<
}
\end{center}

To encrypt a bit $\mu$, choose random $x \in \{0, 1\}^m$, compute ciphertext as follow $(u = Ax, u' = b^tx + \mu\round{\dfrac{q}{2}})$.

To decrypt, compute $u' - s^tu = b^tx +  \mu\round{\dfrac{q}{2}} - s^tAx = (s^tA + e^t)x + \mu\round{\dfrac{q}{2}} - s^tAx = e^tx + \mu\round{\dfrac{q}{2}} \approx \mu\round{\dfrac{q}{2}}$ and test whether it is closer to 0 ($\mu = 0$ case) or $\round{\dfrac{q}{2}} \mod q$ ($\mu = 1$ case).

\subsection{Lattice cryptanalysis of subset sum problem}
LLL algorithm \cite{lll} is a celebrated algorithm, there is a dedicated book just for LLL algorithm and its applications. For the purpose of the article, we'll use only one property of the LLL algorithm. Given a lattice $L(B)$ in $\mathbb{R}^n$ where $B$ is a base of $L$, LLL algorithm will find a relatively short vector $b$: $||b|| < \beta^n \lambda_1(L)$ where $1 < \beta < 2$.

Using the above property, to solve certain equations, we'll prove the following:
\begin{itemize}
\item The solutions belong to a lattice.
\item The solutions are short vectors.
\end{itemize}

After that we'll let LLL algorithm do its job to find short vectors (solutions) in the above lattice for us. Let's apply the described method to solve the following subset sum problem \cite{lllsubsetsum}: given $n$ positive numbers $a_i$ and a positive number $M$, find $x_i \in \{0, 1\}$ such that $a_1x_1 + \cdots + a_nx_n = M$.
\\Let's consider the following vectors
\begin{flalign*}
b_1 &= (1, 0, \cdots, 0, -a_1) & \\
b_2 &= (0, 1, \cdots, 0, -a_2) \\
\vdots \\
b_n &= (0, 0, \cdots, 1, -a_n) \\
b_{n + 1} &= (0, 0, \cdots, 0, M)
\end{flalign*}
\\We have 
\begin{flalign*} 
& x_1b_1 + x_2b_2 + \cdots + x_nb_n + 1b_{n + 1} & \\
&= (x_1, 0, \cdots, 0, -a_1x_1)  \\
&+ (0, x_2, \cdots, 0, -a_2x_2) \\ 
&\vdots \\  
&+ (0, 0, \cdots, x_n, -a_nx_n) \\
&+ (0, 0, \cdots, 0, M)  \\
&= (x_1, x_2, \cdots, x_n, -a_1x_1 - a_2x_2 + \cdots - a_nx_n + M)  \\
&= (x_1, x_2, \cdots, x_n, 0)
\end{flalign*}

I.e., $(x_1, x_2, \cdots, x_n, 0)$ is a vector in the lattice with basis vectors $b_1, \cdots, b_{n + 1}$. Furthermore, $(x_1, x_2, \cdots, x_n, 0)$ is a short vector because $x_i \in \{0, 1\}$. Therefore, LLL algorithm will help find the short vector (solution) $(x_1, x_2, \cdots, x_n, 0)$ for us.

\subsection{Lattice-based homomorphic encryption}
Let's assume we have data that we store in a cloud. To protect our data, we encrypt them and keep
the key to ourselves. On the other hand, we want to take advantage of cloud's computing power, so we
want the cloud to compute on our ciphertexts without knowing what our plaintexts are. Homomorphic
encryption is a special type of encryption that achieves the previous goal. In this section, we'll
describe a simple lattice-based homomorphic encryption \cite{shellencryption}.

We'll use $\ZZ_q[x]/(x^{2^k} + 1)$ ($q$ is a prime number), i.e., polynomials whose coefficients are
in $\ZZ_q$ and all operations are $\mod x^{2^k} + 1$. We also use a small modulus $t$ that is much
smaller than $q$. Note that everything in this section including secret key, message, ciphertext
are polynomials.

The secret key is a polynomial $s$ in $\ZZ_q[x]/(x^{2^k} + 1)$.

To encrypt a message polynomial $m \in \ZZ_t[x]/(x^{2^k} + 1)$ , we randomly generate polynomial
$a$, small error polynomial $e$ and the ciphertext is simply $c = Enc(m) = (c_0, c_1) = (-a, as + m
+ et)$.

To decrypt a ciphertext $c = (c_0, c_1)$, we compute $c_1 + c_0s \mod t = as + m + et - as \mod t =
m + et \mod t = m$.

To see how this encryption is additive homomorphic, let's take a look at two encryptions of $m$ and
$m'$: $c = Enc(m) = (c_0, c_1) = (-a, as + m + et)$ and $c' = Enc(m') = (c_0', c_1') = (-a', a's +
m' + e't)$. If we add $(c_0, c_1)$ and $(c_0', c_1')$ together, we have:

\begin{align*}
(c_0, c_1) + (c_0', c_1') &= (c_0 + c_0', c_1 + c_1') \\
&= (-a - a', as + m + et + a's + m' + e't) \\
&= (-(a + a'), (a + a')s + (m + m') + (e + e')t \\
\end{align*}

If we denote $a'' = a + a'$, $m'' = m + m'$, $e'' = e + e'$, then we see that $c'' = (c_0'', c_1'')
= (c_0, c_1) + (c_0', c_1')$ is the encryption of $m'' = m + m'$ with error $e'' = e + e'$.  To
recap, what we've done is to add 2 ciphertexts together without knowing the messages, but the result
corresponds to the sum of the messages.  It's pretty cool, right?

Another nice property is that if you multiply a polynomial $p$ to the encryption of $m$ then the
result corresponds to encryption of $pm$. To see why it's the case, let's take a look at polynomial
$p$ and encryption of $m$: $(c_0, c_1) = (-a, as + m + et)$. We have:

\begin{align*}
p(c_0, c_1) &= (pc_0, pc_1) \\
&= (p(-a), p(as + m + et)) \\
&=(-pa, pas + pm + pet) \\
\end{align*}

If we denote $a' = -pa$, $m' = pm$, $e' = pe$, then we see that $(c_0', c_1') = p(c_0, c_1)$ is the
encryption of $m' = pm$ with error $e' = pe$.

The final note is that the error increases in both cases. For lattice-based cryptography to work,
the error must be small. Therefore, various techniques have been designed to reduce the error over
time. We won't discuss error reduction techniques here, instead, we'll take a look at an awesome
application of the previous homomorphic encryption in the next section.

\subsection{Lattice-based private information retrieval}
Let's say a server has a public database (e.g. movies, songs, lyrics, stories, books) with $n$ items
$x_1,\cdots,x_n$. A user wants to see a single item $x_i$ at index $i$ from the database without
revealing to the server what item has been downloaded. This is to protect the user's privacy. An obvious
solution is the user downloads all $n$ items from the database. This has perfect privacy, but it
costs significant bandwidth and user's local storage. We'll trade CPU with bandwidth and storage
using the above homomorphic encryption \cite{xpir}. The basic protocol works as follows.

The user forms a sequence of $0$ and $1$ where only at index $i$, it's $1$ while the remaining
numbers are $0$: $0,\cdots, 0, \underbrace{1}_\text{index i}, 0,\cdots,0$. The user uses homomorphic
encryption to encrypt the above sequence, i.e., $c_1 = Enc(0), \cdots, c_{i - 1} = Enc(0),c_i =
Enc(1), c_{i + 1} = Enc(0),\cdots, c_n = Enc(0)$. The user sends $c_1,\cdots,c_n$ to the server.

The server computes $x = x_1c_1 + \cdots + x_nc_n$ without knowing what $i$ is and sends $x$ to the
user.

The user decrypts $x$ and the result is $x_i$. Why's that? By homomorphic property, $x = x_1c_1 +
\cdots + x_nc_n$ corresponds to the encryption of $x_1.0 + \cdots + x_{i - 1}.0 + x_i.1 + x_{i +
1}.0 + \cdots + x_n. 1 = 0 + \cdots + 0 + x_i + 0 + \cdots + 0 = x_i$.

\section{Hash-based signatures}

I won't pretend that I can write better than Matthew Green's \cite{matthewgreenhashbased}, Adam Langley's \cite{adamlangleyhashbased} excellent blog posts about hash-based signatures, so go there and read them :) Anyway, I'll briefly describe the hash-based signatures schemes based on Dan Boneh and Victor Shoup's book \cite{cryptobook} to make this article self-contained.

Hash-based signature scheme is the safest signature scheme against quantum computers. All signature schemes use hash functions and hence they must rely on hash functions' security. All signature schemes, except hash-based signature schemes, must rely on additional security assumptions of other computational hard problems. Hash-based signature scheme, on the other hand, only depends on the security of hash functions. Furthermore, cryptographic hash functions are assumed to act like random oracles and as far as I know, quantum computers have limited success on breaking unstructured functions like random oracles and the best known quantum attack only has quadratic speedup compared to classical attacks. In terms of applied cryptography, I recommend you never deploy stateful hash-based signatures, instead use stateless hash-based signatures although the latter have lower performance. The chance that you screw up security of stateful hash-based signatures deployment is far higher than the chance of real general purpose quantum computers breaking your ECDSA signatures.

If you're familiar with how RSA or ECDSA signatures work and you look at hash-based signatures, you'll find the ideas behind hash-based signatures schemes pretty strange. They're elegant ideas, but they have nothing in common with our familiar RSA and ECDSA signatures. Let's review a few cryptographic concepts that are used as building blocks in hash-based signatures. 

We call a function $f$ a one-way function if given $y = f(x)$, it's hard to find $x$. In practice, standard cryptographic hash functions such as SHA256, SHA-3 are considered one-way functions.

A secure pseudo-random generator (PRG) is a deterministic algorithm $G$ that given a short seed $s$, computes a long output $r$ that is indistinguishable from random sequence. In practice, whenever we call $/dev/urandom$, we actually call a secure PRG function implemented in the operating system. When the operating system boots up, it only collects a small amount of random entropy (e.g. 256 bits) and uses it as seed to compute an arbitrary long stream of random sequence for us. If it's the 1st time you deal with PRG, you would feel that the deterministic property of $G$ somewhat contradicts with the output's randomness property. We can intuitively explain this seemingly paradox as follows. The short seed is random and to the observer (attacker) of the output sequence, the seed is secret. Therefore, while the algorithm $G$ is a deterministic process, as the starting point (seed) is a random secret that no one knows, the whole process to compute output looks like a random process.

A pseudo-random function (PRF) is a deterministic algorithm that takes 2 inputs: a secret key $k$ and data $x$. The security property of PRF is that if the secret key $k$ is chosen at random then the function $f(k,\cdot)$ acts like a random function. In practice, our familiar HMAC(k, x) is a PRF.

We're ready to design hash-based signature schemes. The plan is to construct a one-time signature scheme that is safe to sign a single message. We then extend it to create a q-indexed signature scheme that can securely sign q messages. Finally, we'll build a stateless hash-based signature scheme that can sign practically arbitrary amounts of messages.

\subsection{Lamport's one-time signature}
To sign a 1-bit message $m \in \{0, 1\}$ using one-way function $f$ such as SHA256, Lamport \cite{lamport} invented the following algorithm. The private key $sk$ is 2 large random numbers (e.g. 256-bit) $(x_0, x_1)$. The public key $pk$ is $(y_0, y_1) = (f(x_0), f(x_1))$. The signature $\sigma$ of message $m$ is $\sigma = S(sk, m) = x_m$. To verify the signature of $m$, the verifier checks whether $f(\sigma) \stackrel{?}{=} y_m$.

As $f$ is a one-way function, given the public key $pk = (f(x_0), f(x_1))$, it's hard to compute the the private key $sk = (x_0, x_1)$, so why $(x_0, x_1)$ must be both large and random? If $sk = (x_0, x_1)$ is small or predictable then the attacker can brute-force $x$, compute $f(x)$ and stop until it matches the public key $pk = (f(x_0), f(x_1))$.

It's simple to extend the algorithm to sign a $v$-bit message $m = m_1, \cdots, m_v$. The private key is $(x_{i, 0}, x_{i, 1}), i = \overline{1, v}$. The public key is $(y_{i, 0}, y_{i, 1}) = (f(x_{i, 0}), f(x_{i, 1})), i = \overline{1, v}$. The signature $\sigma$ of message $m$ is $(x_{1, m_1}, \cdots, x_{v, m_v})$.

Why is this algorithm called one-time signature? Let's take a look at an example. Let's say $v = 2$, the private key is $sk = (x_{1, 0}, x_{2, 0}, x_{1, 1}, x_{2, 1})$, the public key is $pk = (f(x_{1, 0}), f(x_{2, 0})$$, f(x_{1, 1}), f(x_{2, 1}))$. The signature of message $m = 00$ is $(x_{1, \bf{0}}, x_{2, \bf{0}})$. If we sign the 2nd message $m = 11$ with signature $(x_{1, \bf{1}}, x_{2, \bf{1}})$ then from the above two signatures, the attacker knows all 4 secret keys $(x_{1, 0}, x_{2, 0}, x_{1, 1}, x_{2, 1})$, i.e., the attacker can itself compute valid signatures of $m = 01$ and $m = 10$. Therefore, the above algorithm can only safely sign a single message.

\subsection{q-indexed signature}
Assuming that we have a one-time signature scheme with key pair $(pk, sk)$ that can sign a single message $m \in M$. If we have $q$ key pairs $(pk_1, sk_1), \cdots, (pk_q, sk_q)$ then we can sign q-messages of the form $(u, m_u) \in \{1, 2, \cdots q\} \times M$. Why? We use key $sk_1$ to sign the message $(1, m_1)$, key $(pk_2, sk_2)$ to sign the message $(2, m_2)$, etc. In other words, the index $u$ decides that we'll use the key $sk_u$ to sign the message $(u, m_u)$.

\subsection{From q-index signatures to stateless many-time signatures}
Let's say we have 2-index signature schemes with key $sk_0$ that can sign 2 messages of the form $(u, m_u), u = 1, 2$. How can we extend it to sign 4 messages? The trick is we use the key $sk_0$ to sign 2 public keys $(pk', pk'')$ and we use the key $(sk', pk')$ to sign 2 messages $(1, m'_1), (2, m'_2)$ and the key $(sk'', pk'')$ to sign 2 messages $(1, m''_1), (2, m''_2)$. Hence we can sign a total of $2^2 = 4$ messages. The structure looks like a Merkle tree where the root is the key $sk_0$ which signs the left child public key $pk'$ and the right child public key $pk''$. These 2 intermediate nodes' keys $(sk', pk')$ and $(sk'', pk'')$ ,in turn, can sign leaf nodes $(1, m'_1), (2, m'_2)$ and $(3, m''_1), (4, m''_2)$. Note that I purposely change the index of $m''$ from $\{1, 2\}$ to $\{3, 4\}$, it doesn't change the fact that it's a 2-indexed signature but the number $3, 4$ reflects the order of leaf nodes from left to right in the tree.

In theory, it's not difficult to extend the above algorithm to sign $2^d$ leaf nodes using a tree of depth $d$. However, there are 2 main problems that have to be solved. The 1st problem is we have to remember which leaf indices that have been used to sign the message. If we reuse the same leaf index to sign 2 different messages then the algorithm becomes insecure because 1 leaf index corresponds to a one-time signature. The 2nd problem is that it's infeasible to generate and store $2^d - 1$ root and intermediate keys.

To solve the 1st problem, we can choose large $d = 256$ and to sign a message $m$, we choose a random 256-bit leaf index. The chance of indices's collision is negligible, so we don't have to remember which indices have been used.

To solve the 2nd problem, we use pseudo-random generator (PRG) and pseudo-random function (PRF) to deterministically generate intermediate nodes's keys on the fly. Let's say the root key is $sk = (k, sk_0)$ where $k$ is our secret seed. To sign a message $m$, we choose a random index leaf node $a = (a_1, \cdots, a_d)$. The index $a = (a_1, \cdots, a_d)$ means that in the path from root to the leaf, we've gone through the following intermediate nodes $(a_1)$, $(a_1, a_2)$, $(a_1, a_2, a_3)$, $\cdots$, $(a_1, a_2, \cdots, a_d)$ where we go to the left if $a_k = 0$ and go to the right  if $a_k = 1$. To compute the key at intermediate node $(a_1, \cdots, a_i)$, we computes a seed for that node $r_i = PRF(k, (a_1, \cdots, a_i))$ and uses the seed $r_i$ to generate the key pairs $(pk_i, sk_i) = PRG(r_i)$ and we sign the generated public key using the key above it in the tree $S(sk_{i - 1}, (a_i, pk_i))$. Finally, we use the key at depth $d - 1$ to sign the message $m$ at depth $d$: $S(sk_{d - 1}, (a_d, m))$.

\section{McEliece and Niederreiter's code-based cryptosystem}
McEliece code-based cryptosystem \cite{mceliece}, \cite{codebasedbook}, \cite{codebasedtalk}, \cite{mcelieceattack} has resisted 4 decades of cryptanalysis. Its security is based on the hardness of certain error-correcting problems which we will study in the following section. It's not popular because its public key is large, but its encryption and decryption are fast. Depending on the context of cryptographic protocols, this limitation may or may not be an issue.

\subsection{Error correcting code}
Let's say we want to transfer a message $m$ of $k = 1$ bit length over the wire. The message is 0 or 1. However, there may be an error in the transmission channel which might cause the bit to be flipped. The problem is how to detect whether there was an error? Another related problem is how can we correct the error?

If we only send the message then there is no way to solve the above problems. We have to send extra bits to help detect the error or to correct it. Instead of sending the bit 0 (correspondingly 1), we send $00$ (correspondingly $11$) and if the transmission channel only has at most $t = 1$ bit of error then we can detect it. Why? If the sender sends $00$ and $t = 1$ bit of error happens then receiver will receive $10$ (if the 1st bit is flipped) or $01$ (if the 2nd bit is flipped). Similar situation happens if the sender sends $11$. I.e., if the receiver receives $01$ or $10$, the receivers knows that there was an error and if the receiver receives $00$ or $11$ it knows that there was no error. This solves the problem of error detection, but it doesn't solve the problem of error correction problem. It's not hard to convince ourselves that if we encode $0$ as $000$ and $1$ as $111$ and if the transmission channel has at most $t=1$ bit errors then we can correct the flipped bit.

Let's pause for a moment to introduce the terminologies. The sender wants to send a message $m$ of length $k$ (e.g. $0$ or $1$). The sender encodes the message into a different form called codeword $c$ of length $n, n > k$ (e.g. $000$ or $111$). The transmission channel may introduce $t$ bit error $e$ into $c$, i.e., the receiver receives $\hat{c} = c + e$ where $e$ has at most $t$ bits 1. The receiver decodes $t$ bit error from $\hat{c}$ to get $c$ and deduce $m$ from $c$. If the receiver can correctly fix up to $t$ bit errors then we say we have t-error correcting code.

In this article, we're only concerned with binary (aka $F_2$) linear code, i.e., we have a generator matrix of size $k \times n$, each matrix entry is in $F_2$  and the codeword is $c = mG$. A binary message $m$ of size $1 \times k$ will produce a codeword of size $1 \times n$. Why is it called linear code? It's because $mG$ is the linear combination of rows of matrix $G$. In this case, we call the set $C$ of all codewords $c$ a $(n, k)$-code. A matrix $H$ of size $(n - k) \times n$ whose null space is $C$ (i.e., $Hc^t = 0$) is called parity-check matrix. Note that the kernel equation $Hc^t = 0$ is an alternative way to define codes besides using generator matrix. Let's see why $H$ is called check matrix. We have $H\hat{c}^t = H(c + e)^t = Hc^t + He^t = 0 + He^t = He^t$, i.e., if there was no error ($e = 0$) then $H\hat{c}^t = 0$. The value $H\hat{c}^t = He^t$ is called the syndrome of $\hat{c}$. 

\subsection{Goppa codes}
McEliece original cryptosystem uses binary Goppa codes \cite{goppa} and it has resisted cryptanalysis for 4 decades. While it's possible to replace binary Goppa codes with other error-correcting codes, a few proposals using alternative codes have been broken over time. Therefore, in this section, we briefly introduce binary Goppa codes.

We will work in finite field $F_{2^m}$. Fix a list of $n$ elements $a_1, a_2, \cdots, a_n$ in $F_{2^m}$. Choose a degree-t irreducible polynomial $g(x) \in F_{2^m}[x]$. As $g(x)$ is irreducible, $F_{2^m}[x]/g(x)$ (i.e. polynomial where coefficients are in $F_{2^m}$ and all operations are $\mod g(x)$) forms a finite field and hence every element has an inverse. The binary Goppa code is defined as follow using kernel equation 
\begin{center}
$\Gamma(a_1, a_2, \cdots, a_n, g) = \{ c \in F_2^n: \sum_{i} \frac{c_i}{x - a_i} = 0 \mod g(x)\}$
\end{center}

I guess you're confused because in the previous section, we use matrix notation to define error-correcting code, but here we use polynomial. What is the relationship between polynomial form and matrix form?

Note that $\frac{1}{x - a_i}$ is a shortcut notation to denote the inverse of $x - a_i$ in the finite field $F_{2^m}[x]/g(x)$. There is an efficient algorithm to compute the inverse of $x - a_i$ and let's denote $g_{i} = \sum_{j = 0}^{t - 1} g_{i,j}x^j$ the inverse of $x - a_i$. Now, rewrite the kernel equation $\sum_{i} \frac{c_i}{x - a_i} = 0 \mod g(x)$ as follow
\begin{align*}
\sum_{i = 1}^n c_i\frac{1}{x - a_i} &= \sum_{i = 1}^{n} c_i\sum_{j = 0}^{t - 1} g_{i,j}x^j \\
        &= \sum_{j = 0}^{t - 1} \bigg( \sum_{i = 1}^n g_{i,j} c_i \bigg) x^j \\
        &= 0 \mod g(x)
\end{align*}

The last equation means that $x^j$'s coefficient $\sum_{i = 1}^n g_{i,j} c_i$ must be zero for all $j = 0, \cdots, t - 1$. I.e., we have a system of equations $\sum_{i = 1}^n g_{i,j} c_i = 0, j = 0, \cdots, t - 1$ where $c = (c_1, c_2, \cdots, c_n)$ is our codeword and that is our familiar matrix form of kernel equation. 
         
\subsection{McEliece's cryptosystem}
McEliece's cryptosystem is based on the following observation. In binary Goppa codes, if the receiver/decoder knows the generator matrix $G$ then it's easy to decode and correct errors in the transmission channel. However, it's difficult to solve the decoding problem for arbitrary $(n,k)$ linear code. Therefore, McEliece cryptosystem generates a generator matrix, keeps it as private key and scrambles it to make it look random and use the scrambled version as the public key.

In details, the sender chooses $k \times n$ Goppa generator matrix $G$ that can correct up to $t$ errors, $k \times k$ binary non-singular matrix $S$, $n \times n$ permutation matrix $P$. The matrices $S, P$ are used to hide the generator matrix $G$.

The public key is $G' = SGP$, the private key is $(S, G, P)$.

To encrypt a message $m$ of length $k$, choose a random error vector $e$ that has $t$ bits 1, compute the ciphertext $c = mG' + e$.

To decrypt $c$, compute $cP^{-1} = (mG' + e)P^{-1} = (mSGP + e)P^{-1} = (mS)G + eP^{-1}$. As $eP^{-1}$ is just a permutation of $e$, it has $t$ bits 1. Therefore, with the knowledge of $G$, the receiver can use an efficient decoding algorithm to deduce $mS$. To recover $m$, compute $(mS)S^{-1} = m$.

\subsection{Niederreiter's cryptosystem}
Niederreiter's \cite{niederreiter}, \cite{mcelieceattack} cryptosystem is a variant of McEliece cryptosystem where it has the same security as McEliece cryptosystem. Recall that to define error-correcting code, we can either use generator matrix or use parity-check matrix. McEliece cryptosystem uses generator matrix while Niederreiter cryptosystem uses parity-check matrix.

In details, the sender chooses a $(n - k) \times n$ parity-check matrix $H$ of Goppa codes that can correct up to $t$ errors, a $(n - k) \times (n - k)$ binary non-singular matrix $S$, $n \times n$ permutation matrix P. The matrices $S, P$ are used to hide the parity check matrix $H$.

The public key is $K = SHP$, the private key is $(S, H, P)$.

To encrypt a message $m$ of length $n$ and has $t$ bits 1, compute the ciphertext $c = Km^t$.

To decrypt, compute $S^{-1}c = S^{-1}SHPm^t = HPm^t = H(Pm^t)$. $Pm^t$ is just a permutation of $m$ so it has $t$ bits 1. With the knowledge of parity check matrix $H$, the receiver can use an efficient decoding algorithm to compute $Pm^t$ and hence $m^t = P^{-1}Pm^t$.

\bibliographystyle{unsrt}
\bibliography{research}
\end{document}